\documentclass[%
reprint,
showpacs,preprintnumbers,showkeys,
footinbib,
 amsmath,amssymb,
 aps,
pra,
superscriptaddress
]{revtex4-1}

\usepackage{graphicx}% Include figure files
\usepackage{dcolumn}% Align table columns on decimal point
\usepackage{bm}% bold math
\usepackage{makecell}% multiple lines in a cell
\usepackage{multirow}% combine rows in tables
\begin{document}

\title{Condensation versus Long-range Interaction: Competing Quantum Phases in Bosonic Optical Lattice Systems at Near-resonant Rydberg Dressing}

\author{Andreas Gei{\ss}ler}
\email{geissler@th.physik.uni-frankfurt.de}
\affiliation{Institut f\"ur Theoretische Physik, Goethe-Universit\"at, 60438 Frankfurt/Main, Germany}
\author{Ivana Vasi\'c}
\affiliation{Scientific Computing Laboratory, Institute of Physics Belgrade, University of Belgrade, Pregrevica 118, 11080 Belgrade, Serbia}
\affiliation{Institut f\"ur Theoretische Physik, Goethe-Universit\"at, 60438 Frankfurt/Main, Germany}
\author{Walter Hofstetter}
\affiliation{Institut f\"ur Theoretische Physik, Goethe-Universit\"at, 60438 Frankfurt/Main, Germany}

\date{\today}

\begin{abstract}
Recent experiments have shown that (quasi-)crystalline phases of Rydberg-dressed quantum many-body systems in optical lattices (OL) are within reach. Rydberg systems naturally possess strong long-range interactions due to the large polarizability of Rydberg atoms. Thus a wide range of quantum phases have been predicted, such as a devil's staircase of lattice incommensurate density wave phases as well as more exotic lattice supersolid order for bosonic systems, as considered in our work. Guided by results in the ``frozen'' gas limit, we study the ground state phase diagram at finite hopping amplitudes and in the vicinity of resonant Rydberg driving, while fully including the long-range tail of the van der Waals interaction. Simulations within real-space bosonic dynamical mean-field theory (RB-DMFT) yield an extension of the devil's staircase into the supersolid regime where the competition of condensation and interaction leads to a sequence of crystalline phases.
\end{abstract}

%Use showkeys class option if keyword

\maketitle

%\tableofcontents

Despite the high tunability of ultracold atomic systems as analog quantum emulators, strong long-range correlations still represent an important challenge in the field. While Feshbach resonances give access to tunable local interactions \cite{Bloch2008a}, recent experimental breakthroughs allow for the trapping, cooling and control of  ultracold polar molecules, as well as magnetic \cite{Lahaye2007} and Rydberg atoms \cite{Baranov2012}. The significance of Rydberg excitations for creating strong non-local correlations has been pointed out \cite{Saffman2010, Weimer2008, Honer2010}.

Recent experiments have studied the statistical properties of dissipative Rydberg gases \cite{Malossi2014, Schempp2014} and especially of superatoms \cite{Niederprum2015, Zeiher2015, Weber2014b}, where the Rydberg blockade effect was analyzed. Using electromagnetically induced transparency, the occurrence of diffusive F\"orster energy transport has been shown \cite{Gunter2013a}. Also, ultralong-range Rydberg molecule formation has already been observed \cite{Bendkowsky2009}, while crystallization of Rydberg atoms has been achieved up to a small number of excitations in the ``frozen'' limit \cite{Schauß2012a, Schauß2014}. There the system behaves like a spin-$\frac{1}{2}$ model with imbalanced interactions, as analyzed in numerous theoretical works \cite{Vermersch2014a, Pohl2010a, Schauß2014, Lesanovsky2014a, Schachenmayer2010, Capogrosso-Sansone2009b}, predicting a series of lattice incommensurate ordered phases (``devil's staircase''). The opposite limit of weak Rydberg dressing has extensively been investigated in theory \cite{Johnson2010, Honer2010, Pupillo2010, Cinti2010, Wuster2011, Li2011a}, predicting the formation of (droplet) supersolids (SS), while its experimental realization remains an open challenge \cite{Balewski2013a, Niederprum2015, Jau2015, Zeiher2016}. 

\section{System}\label{sec:Sys}
In this work we focus on the far less understood intermediate regime of finite hopping at near-resonant and coherent excitation of the Rydberg state. Previous work in this regime so far only considered either the nearest-neighbour (NN) limit for the interactions in a Gutzwiller mean-field simulation \cite{Saha2014}, or the low-dimensional case \cite{Sela2011} with vanishing single particle hopping \cite{Weimer2010}. In the following we will introduce our approach for obtaining a ground state phase diagram. The combination of a ``frozen'' limit model and a real-space extension of bosonic dynamical mean-field theory (RB-DMFT) allows for an efficient quantitative analysis of the phase diagram for arbitrary range of the interaction. We will first introduce the two-species ``frozen'' limit model, which we solve in the Hartree-approximation. Then we outline the calculation of the phase diagram using RB-DMFT. Finally we discuss the obtained quantum phases and the different types of long-range order observed.

Considering both ground $\left| g \right\rangle$ and Rydberg excited $\left| e \right\rangle$ states, our full grand canonical Hamiltonian (in natural units $\hbar = 1$) can be written in terms of bosonic annihilation operators $\hat{b}_{\sigma,i}$ acting on site $i$ of a square optical lattice (OL), where $\hat{n}^{\sigma}_{i} = \hat{b}^{\dagger}_{\sigma,i}\hat{b}_{\sigma,i}$ and $\sigma=g,e$:

\abovedisplayskip=-5pt %arxiv-version only

\begin{align}
H = H_{2BH,kin} + \sum_i^N \left( H_{2BH,loc,i} + H_{R,i} + H_{vdW,i} \right)
\label{eq:full-hamil}
\end{align}
with the kinetic energy given by hopping of strength $J$ and $\eta J$ between all pairs of nearest neighbours (NN) $\langle i,j \rangle$ as $H_{2BH,kin} = -J\sum_{\langle i,j \rangle} \left( \hat{b}^{\dagger}_{g,i} \hat{b}_{g,j} + \eta \hat{b}^{\dagger}_{e,i} \hat{b}_{e,j} + h.c. \right)$ and local interaction terms for a two species model included in 

\begin{align}
H_{2BH,loc,i} =& U \left( \frac{\hat{n}_i^g}{2} (\hat{n}_i^g-1) + \lambda \hat{n}_i^g \hat{n}_i^e + \tilde{\lambda} \frac{\hat{n}_i^e}{2} (\hat{n}_i^e-1) \right) \nonumber \\ &-\mu \left( \hat{n}_i^g + \hat{n}_i^e \right)   \label{eq:local-hamil}
\end{align}
where $U,\lambda U$ and $\tilde{\lambda} U$ are the strengths of the three Hubbard interaction terms and $\mu$ is the chemical potential. The excited electronic (Rydberg) states of the atoms are populated via coherent driving, which leads to Rabi oscillations. This process is induced by the interaction with the laser light field (see for example equation (A.11) in Chapter V of \cite{Cohen2004}). So, when using the interaction picture, for a given atom we have

\begin{align}\label{eq:Inter-hamil}
H_R^{(I)} =& -\hat{\mathbf{d}} \cdot \mathbf{E}_0 \cos(\omega_L t) \\ =& \frac{\Omega}{2} \left(e^{-i\omega_L t} + e^{i\omega_L t} \right) \hat{\sigma}^+(t) \nonumber \\ &+ \frac{\Omega}{2}^* \left(e^{-i\omega_L t} + e^{i\omega_L t} \right) \hat{\sigma}^-(t)\nonumber
\end{align}
The time dependence of the (pseudospin-flip) $\hat{\sigma}^{\pm}$-operators is given by the transition frequency $\omega_0$, while $\omega_L$ is the frequency of the light field. If we thus insert $\hat{\sigma}^{\pm}(t) = \hat{\sigma}^{\pm}_0 e^{\pm i \omega_0 t}$ into (\ref{eq:Inter-hamil}), while also replacing $\hat{\sigma}^{\pm}_0$ by appropriate products of bosonic creation and annihilation operators, we obtain the full expression in the interaction picture.

\begin{align}\label{eq:I-Rabi-hamil}
H_R^{(I)} =& \frac{\Omega}{2} \left(e^{-i\Delta t} + e^{i(\omega_L + \omega_0)t} \right) \hat{b}^{\dagger}_e \hat{b}_g \\ &+ \frac{\Omega}{2}^* \left(e^{-i(\omega_L + \omega_0) t} + e^{i\Delta t} \right) \hat{b}^{\dagger}_g \hat{b}_e \nonumber
\end{align}
We can then assume that the Rabi frequency $\Omega$, as given by the dipole moment of the transition and the strength of the light field, is a real quantity. The detuning $\Delta = \omega_L - \omega_0$ defines the slow time scale. Terms oscillating with fast frequencies can be discarded if $\Delta \ll \omega_L + \omega_0$, yielding the rotating wave approximation \cite{Robicheaux2005}. The time-independent Hamiltonian in the rotating wave approximation follows from the unitary transformation, defined by the time-dependent unitary transformation matrix $U=U(t) = \hat{b}^{\dagger}_g\hat{b}_g + e^{i\Delta t} \hat{b}^{\dagger}_e\hat{b}_e$:

\begin{align}
H_R = U H_R^{(I)} U^{-1} + i\frac{dU}{dt}U^{-1} = \frac{\Omega}{2} \left( \hat{b}^{\dagger}_g\hat{b}_e + \hat{b}^{\dagger}_e\hat{b}_g \right) - \Delta \hat{n}^e
\end{align}
This follows straight from $[U,\frac{d}{dt}] = - (\frac{dU}{dt})$, which simply has to be inserted into the Schr\"odinger equation, while the wave function transforms as $\tilde{\psi} = U \psi$. Thus the Rabi process for each lattice site in the rotating wave approximation takes the following form.

\begin{align}\label{eq:Rabi-hamil}
H_{R,i} = \frac{\Omega}{2} \left( \hat{b}^{\dagger}_{g,i}\hat{b}_{e,i} + \hat{b}^{\dagger}_{e,i}\hat{b}_{g,i} \right) - \Delta \hat{n}^e_i
\end{align}

In addition we also consider the non-local van der Waals (vdW) interaction between Rydberg states. At distances relevant in OLs, it is dominated by its long-range tail, thus for atoms at sites $\mathbf{i}$ and $\mathbf{j}$

\begin{align}\label{eq:bare-vdW}
H_{vdW,i} = \frac{V_{vdW}}{2} \sum_{j \neq i} \frac{\hat{n}^e_i \hat{n}^e_j}{\left| \mathbf{i}-\mathbf{j} \right|^6}
\end{align}
where $V_{vdW} = C_6/a^6$ with the vdW coefficient $C_6$ and the lattice parameter $a$. This model has been previously investigated in the limit of NN interactions only, by applying Gutzwiller mean-field theory \cite{Saha2014}. In our study we go beyond this common approximation and show that the phase diagram is far richer.

Many of the above model parameters are easily adjustable in experiments, some even over several orders of magnitude. The Rabi parameters can be directly controlled via the laser intensity (Rabi frequency $\Omega$, which also depends on the matrix elements of the chosen transition) and laser detuning $\Delta$ \cite{Bloch2008a}, while the vdW interaction is determined by the Rydberg level considered.
The remaining parameters are not as simple to control. The hopping of Rydberg-excited atoms is not yet an experimentally well-controlled parameter, since the OL, trapping the ground state (GS) atoms, is not the same for Rydberg states by default. Often it is even of opposite sign \cite{Dutta2000,Younge2010,Younge2010a}. Here we focus on the limiting case $\eta = 0$, motivated by the fact that the Rydberg part of the Hilbert space is dominated by the vdW interaction and the Rabi frequency, while the total kinetic energy contribution from $\left| e \right\rangle$ will be small compared to $\left| g \right\rangle$, due to low Rydberg fractions (similar to \cite{Honer2010}, see Appendix \ref{app:Rydhopp}). As the Rydberg states are perturbed by the Rabi process, their localization will anyway be lifted due to hybridization of $|e\rangle$ with $|g\rangle$.

Local interactions are fixed by considering the Quantum Zeno Effect \cite{Misra1977,Garcia-Ripoll2009,Vidanovic2014a}. It describes the observation that loss channels with a bare loss rate $\gamma_0 \gg U$ are strongly suppressed in the lattice, as this corresponds to a strong measurement of the lossy states, thus keeping them fixed at zero occupation. Experiments have shown the large cross section of molecular ion formation in Rydberg gases \cite{Niederprum2015}. Due to the different electronic structure of such ions, they are not trapped by the confining potential, implying a large bare loss rate $\gamma_0$. The local quantum states susceptible to molecule formation or ionization correspond to Fock states of the form $| n_g \geq 0 , n_e > 0 \rangle$. We can model their loss-induced suppression by choosing ``arbitrary'' large values for both $\lambda$ and $\tilde{\lambda}$ \eqref{eq:local-hamil}.

\section{``Frozen'' limit model}\label{sec:frozen limit}
Due to the many possible spatial crystalline orderings, an efficient method to distinguish them is needed. Therefore we first analyze the ``frozen'' limit, where all spatial hopping terms are zero ($J=0$). This allows for a simple analytical investigation of the ground state manifold with few approximations. Moreover it makes for a useful exact starting point for considering finite hopping ($J \neq 0$), which we simulate within RB-DMFT.
Assuming a mean lattice filling $\bar{n} < 1$, where $\bar{n} = \sum_i ( \langle \hat{n}_i^g \rangle + \langle \hat{n}_i^e \rangle )/N$, only empty or singly occupied sites are to be expected. We may also assume that such a system always has a spatially periodic ground state. For such crystalline order, we consequently only need to consider those sites $i$ of the full Hamiltonian which are non-empty, in order to calculate the energy:

\begin{align}\label{eq:atomic_limit_model}
\begin{aligned}
H_i = & \frac{\Omega}{2} \left( \hat{b}^{\dagger}_{g,i}\hat{b}_{e,i} + \hat{b}^{\dagger}_{e,i}\hat{b}_{g,i} \right) - \Delta \hat{n}^e_i \\ & -\mu \left( \hat{n}_i^g + \hat{n}_i^e \right) + \frac{V_{vdW}}{2} \sum_{j \neq i} \frac{\hat{n}^e_i \hat{n}^e_j}{\left| \mathbf{i}-\mathbf{j} \right|^6}
\end{aligned}
\end{align}

\begin{figure}[h]
 \centering
  \includegraphics[width=0.99\columnwidth]{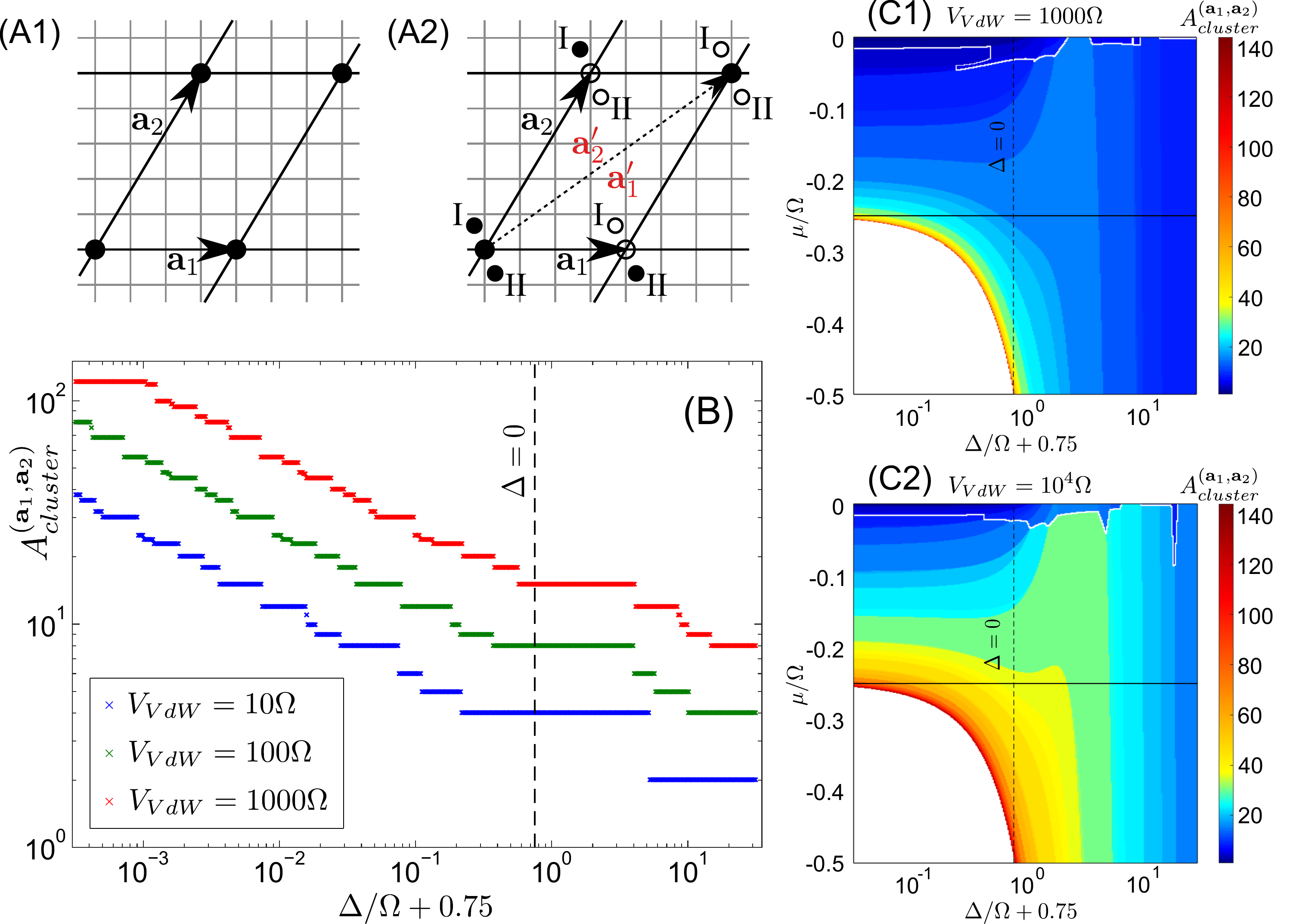}
 \caption{(A1) Spanning vectors $(\mathbf{a}_1,\mathbf{a}_2)$ define the Bravais cell of a superlattice for the underlying OL (gray). Black filled circles correspond to occupied sites, while remaining sites are empty. (A2) Possible chequerboard generalizations of (A1), where spanning vectors connect two different sublattices (filled and empty circles). Mapping to the striped versions (I/II) is explained in the text. In (B,C) different crystalline phases of the ``frozen'' limit model can be distinguished by $A^{(\mathbf{a}_1,\mathbf{a}_2)}_{cluster}$. (B) shows devil's staircases for the logarithmic approach to $\Delta/\Omega = -3/4$ at $\mu_0/\Omega = -1/4$ (solid lines in C1,C2). Phases above white lines (C1,C2) correspond to two-sublattice order with canted state orientation.}
 \label{fig:spanning_vecs}
 \label{fig:atomic_phasediag}
\end{figure}

Any periodic superlattice structure can be constructed from a set of spanning vectors (one per spatial dimension, Fig.~\ref{fig:spanning_vecs}(A1)), which in our case are restricted to the discrete set of points given by the OL. Applying the Hartree approximation for a given set of spanning vectors $(\mathbf{a}_1, \mathbf{a}_2)$, the Hamiltonian reduces to a set of self-consistent single-site problems with at most $A^{(\mathbf{a}_1,\mathbf{a}_2)}_{cluster}$ different self consistent values $n_j^e=\langle \hat{n}_j^e \rangle$, where $A^{(\mathbf{a}_1,\mathbf{a}_2)}_{cluster}$ is the area spanned by the given vectors. Due to low filling $\bar{n} < 1$ we only consider two values ($n_A^e,n_B^e$), where each corresponds to one of the two sublattices defined by their sets of sites $A$/$B$ (indicated by empty/filled circles in Fig.~\ref{fig:spanning_vecs}(A2)) of a chequerboard version of the spanned superlattice.

\begin{figure}[h]
 \centering
  \includegraphics[width=0.8\columnwidth]{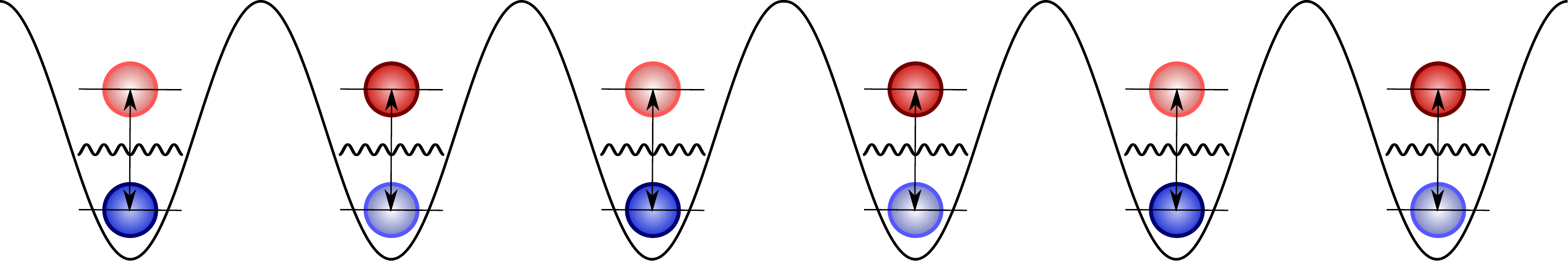} %\newline
 \caption{Schematic representation of a one dimensional CIAF state in an optical lattice. Colored circles correspond to the ground (blue) and excited (red) Fock-states and the opacity is related to the amplitudes in the local linear combinations (\ref{eq:general frozen state}). A complete polarization of the state is suppressed by the Rabi process induced by the incident light field (small black waves and arrows).}
 \label{fig:CIAF}
\end{figure}

For given vectors $(\mathbf{a}_1,\mathbf{a}_2)$ two further versions are indicated by I/II in Fig.~\ref{fig:spanning_vecs}(A2), where one of the two transformations $\mathbf{a}_{1/2} \rightarrow \mathbf{a}_{1/2}' = \mathbf{a}_{1/2} + \mathbf{a}_{2/1}$ was applied. This allows for energy optimization via canted state orientation, which is equivalent to canted Ising antiferromagnetic (CIAF) order and becomes important for increased lattice fillings. Generally, ``frozen'' states (within the Hartree approximation) can be written as

\begin{align} \label{eq:general frozen state}
\left| \Psi \rangle\right. = \prod_{C} \prod_{i \in C}^N \left( \cos \phi_i \left| \downarrow \rangle\right._i + e^{i \theta_i} \sin \phi_i \left| \uparrow \rangle\right._i \right)
\end{align}
where the state of the full system is given by a product over a lattice of unit cells $C$ containing $N$ sites each, with an internal structure given by the set of $\phi_i \in [0,\pi/2]$ and $\theta_i \in [0,2\pi]$ for $i=1,\ldots ,N$. Setting at least one $\phi_i \not\in \{ 0,\pi/2 \}$ yields CIAF order. In case of the Mott-like ``frozen'' limit, the not yet specified quasi-spin states can in principle be any set of two bosonic Fock states, also including the empty vacuum state $| n_g = 0 , n_e = 0 \rangle$. Note that the use of different particle numbers for the states at a site $i$, as for example the combination of an empty site with any allowed Fock state on this site, implies $\phi_i =  0;\pi/2$. Also note that $\theta_i = \pi$ combined with $(\downarrow,\uparrow) = (g,e)$ corresponds to a dark state, as is used for an $s$-state to $s$-state transition (required for isotropically interacting $^{87}$Rb Rydberg states) to suppress decay via the intermediary $p$-state. An example of CIAF order is schematically shown in Fig.~\ref{fig:CIAF}, where the two sublattices correspond to the $A/B$ sites.

For the interaction energy for each sublattice within Hartree approximation we obtain ($A \leftrightarrow B$)

\begin{align}\label{eq:2sub_Hartree}
H_{vdW,A}^{\mathrm{Hartree}} = V_{vdW} \hat{n}_A^e \left( \sum_{\mathbf{j} \in A \setminus 0} \frac{\langle n_A^e \rangle}{\mathbf{j}^6} +\sum_{\mathbf{j} \in B } \frac{\langle n_B^e \rangle}{\mathbf{j}^6} \right)
\end{align}
where $\mathbf{j}$ points from a given site (0) of $A$ to any site of both $A$ and $B$. Thus the site-averaged grand canonical potential $f$ is simply given by $f = \sum_i{\langle H_i \rangle} / A^{(\mathbf{a}_1,\mathbf{a}_2)}_{cluster} = \sum_{i=A,B}\frac{\langle H_i \rangle}{2} \cdot \bar{n}$ with the vdW interaction evaluated by~\eqref{eq:2sub_Hartree}.

\begin{figure*}[t]
 \centering
  \includegraphics[width=0.98\textwidth]{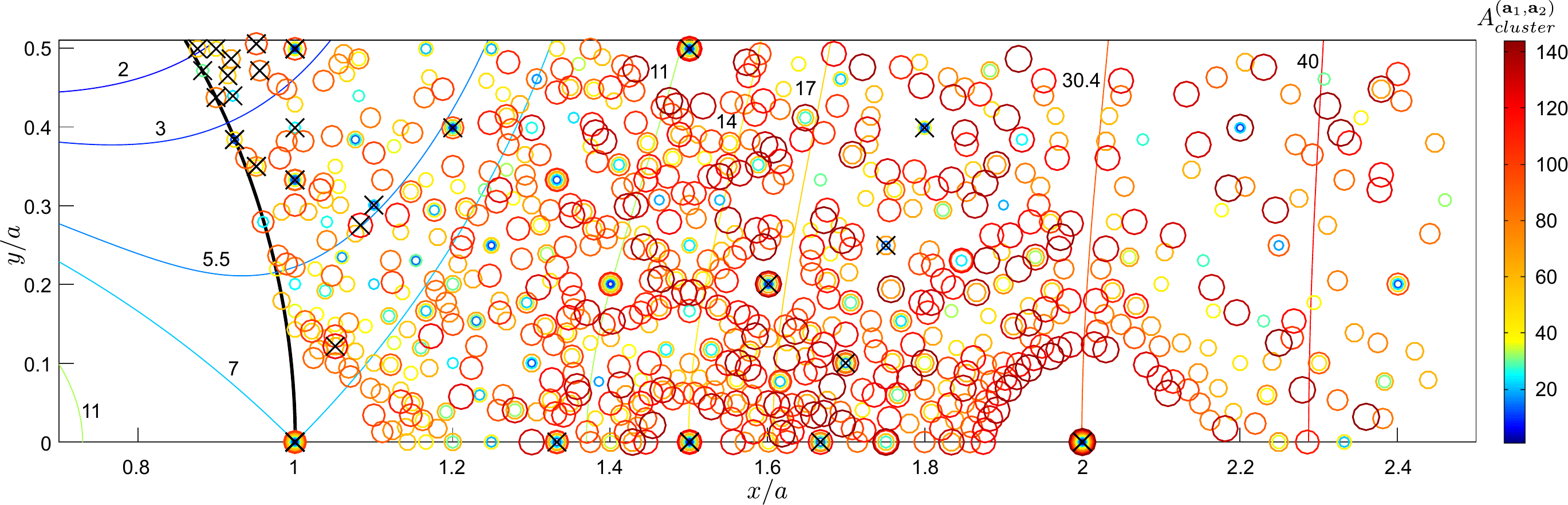}
 \caption{Each colored marker represents a pair of tested spanning vectors from the set $\mathcal{V}_s$. Their coordinates are given by the larger vector, after a combined scaling and rotation of both vectors, so that the smaller vector is mapped onto $(0,1)$. They can thus only appear outside of the unit circle (thick black line). Their color and size corresponds to the area of each crystal unit cell. In addition, also some contour lines for $R_{(0,1)}^{(x,y)}$ are shown. Crossed markers correspond to crystal structures actually appearing as ground states of the atomic limit model in the blue detuned regime for $V_{vdW}<10^4 \Omega$.}
 \label{fig:atomic_limit_configs}
\end{figure*}

Minimizing $f$ with respect to a set $\mathcal{V}_s$ of spanning vectors then yields the many-body ground state phase diagram in the ``frozen'' limit and for $\bar{n} < 1$, as shown in Fig.~\ref{fig:atomic_phasediag}(B,C). For this variational minimization it is useful to represent the remaining sums over the sublattices $A$ and $B$ as functions of the spanning vectors

$$V_{\mathbf{a}_1}^{\mathbf{a}_2} = \sum_{\mathbf{j} \in A \setminus 0} \frac{1}{\mathbf{j}^6}\textrm{ and }
W_{\mathbf{a}_1}^{\mathbf{a}_2} = \sum_{\mathbf{j} \in B} \frac{1}{\mathbf{j}^6}$$
while it is furthermore helpful to introduce 

\begin{align}
R=R_{\mathbf{a}_1}^{\mathbf{a}_2} = \mathrm{max}(V_{\mathbf{a}_1}^{\mathbf{a}_2},W_{\mathbf{a}_1}^{\mathbf{a}_2}) /\mathrm{min}(V_{\mathbf{a}_1}^{\mathbf{a}_2},W_{\mathbf{a}_1}^{\mathbf{a}_2})
\end{align}
as the crystal structure-dependent ratio of the long-range interaction sums. The dependence of $R_{\mathbf{a}_1}^{\mathbf{a}_2}$ on the spanning vectors is shown by the contour lines in Fig.~\ref{fig:atomic_limit_configs}. It should be noted that there is no dependence on the actual form of the interaction, as we use a scale free long-range interaction in the present case.
In order to perform the minimization procedure, we generate a set $\mathcal{V}_s$ (as shown in Fig.~\ref{fig:atomic_limit_configs}), which needs to at least represent the whole range of superlattices, which can in principle be expected in the regime under consideration. In our ``frozen'' model (\ref{eq:atomic_limit_model}) the onsite interaction $U$ is neglected for $\bar{n}<1$. With $\Omega$ as the energy scale, only $V_{vdW}$, $\Delta$ and $\mu$ remain as tunable parameters, defining the region to be investigated.

Especially $V_{vdW}$ is important for the choice of $\mathcal{V}_s$, as it defines the blockade radius $R_b = (C_6/\Omega)^{1/6}$ for Rydberg excitations, which corresponds to a radius of up to 5 OL sites for $V_{vdW} \le 10^4 \Omega$. On a square lattice this would correspond to a volume of up to 25 lattice sites. In order to allow for even lower fillings, enabled by the chemical potential or the detuning, we will consider volumes of up to $12 \times 12$ lattice sites. The complete set $\mathcal{V}_s$ of spanning vectors used here is shown in Fig.~\ref{fig:atomic_limit_configs}, modulo similarity transformations for each pair.

\begin{figure}[h]
 \centering
  \includegraphics[width=0.98\columnwidth]{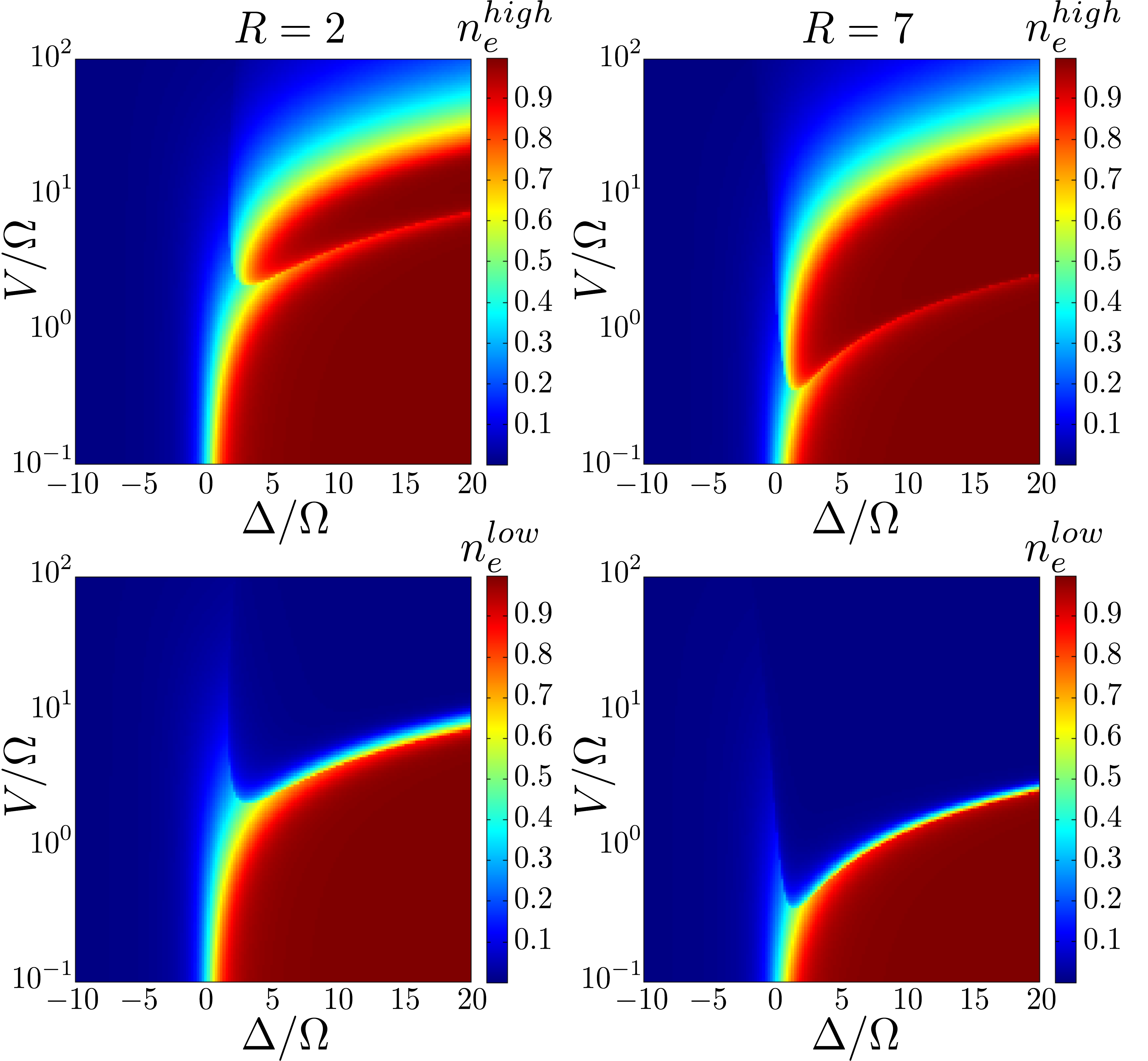}
 \caption{Various solutions of the effective ``frozen'' model (\ref{eq:eff_frozen_rydfrac}) for $R=2,7$. Shown are the Rydberg fractions $n_e$ for each of the two sublattices, which are respectively indexed by whether the sublattice with high or low Rydberg fraction is considered. Canting appears if $n_e^{low} \neq n_e^{high}$ and at least one of them is not equal to unity.}
 \label{fig:effective_fractions}
\end{figure}

If we then also define $V = V_{vdW}\cdot\mathrm{min}(V_{\mathbf{a}_1}^{\mathbf{a}_2},W_{\mathbf{a}_1}^{\mathbf{a}_2})$ and use the Rabi frequency $\Omega$ as energy scale, the self consistency conditions for $n^e_{A/B} = \langle \hat{n}^e_{A/B} \rangle$ in the many-body ground state are given by

\begin{align}
\begin{aligned}
n^e_{A/B} =& 1-n^e_{A/B} \left| V(n^e_{B/A}+R n^e_{A/B}) - \Delta \right.  
 \\ &+ \left.\sqrt{1+\left( V(n^e_{B/A}+R n^e_{A/B}) - \Delta \right)^2} \right| 
\end{aligned} \label{eq:eff_frozen_rydfrac}
\end{align}
The solutions of this effective model, where $R$ is just the the ratio of any inter- and intra-sublattice interactions, are shown in Fig.~\ref{fig:effective_fractions} for some relevant values of $R$ (compare Fig.~\ref{fig:atomic_limit_configs}).

As $f = \sum_{i=A,B}\frac{\langle H_i \rangle}{2} \cdot \bar{n}$ within these limits and approximations, its minimization with respect to our set $\mathcal{V}_s$ yields the many-body ground state phase diagram in the atomic limit and for $\bar{n} < 1$, as is shown in Fig.~\ref{fig:atomic_phasediag}(B,C). In the comparison of all lattice structures from the set $\mathcal{V}_s$, as shown in Fig.~\ref{fig:atomic_limit_configs}, the configurations of minimal energy anywhere in the analyzed parameter region (compare parameter ranges in Fig.~\ref{fig:atomic_phasediag}(B,C)) are marked by a cross. Those points primarily accumulate where they correspond either to triangular order $\mathbf{a}_2 = (\sqrt{3}/2, 1/2)$ or a square lattice $\mathbf{a}_2 = (1,0)$. Points with increased $R>10$ on the other hand are more susceptible to the formation of CIAF order (as can be seen in Fig.~\ref{fig:effective_fractions}). If one then only considers one of the two sublattices, as for example the one with increased Rydberg fraction, it again resembles triangular order more closely, as is possible without the canted order, while keeping the lattice filling constant. On the other hand, no spanning vectors with minimal energy are to be found beyond a radius of 2, especially the point $(2,0)$ is the most distant (see Fig.~\ref{fig:atomic_limit_configs}), which rules out stripe-like order.

From \eqref{eq:atomic_limit_model} in Hartree approximation it furthermore follows that the chemical potential $\mu_0$, determining the transition to the vacuum state, is given by

\begin{align}
\mu_0 = -\frac{\Delta+\sqrt{\Omega^2+\Delta^2}}{2}.
\end{align}
Approaching this limit by varying either $\mu$, $\Delta$ or $\Omega$ yields a devil's staircase of fractional lattice commensurate fillings (see Fig.~\ref{fig:atomic_phasediag}B), stabilized by the long-range interactions. Note that our ansatz only allows for fillings of the form $\frac{1}{n}$ with $n \in \mathbb{N}$ (see also \cite{Lauer2012a, Rademaker2013a}).

\section{Itinerant case within RB-DMFT}\label{sec:DMFT}
We now use the ``frozen'' limit results as an exact starting point for our RB-DMFT simulations, since both models map onto each other in Hartree approximation, for vanishing $J$ and $\bar{n} < 1$. However for non-zero $J$ we cannot expect the crystal symmetry to always be given by the ``frozen'' limit results. Therefore also other crystalline structures corresponding to similar mean inter-atom distances are simulated. Furthermore RB-DMFT requires a truncation of the local Fock-space. Since a hard cutoff, using only the first $N_c$ Fock states, strongly restricts the maximum observable local particle number in a condensate, we instead use a soft cutoff utilizing the \textit{coherent-tail} state $\propto \sum_{n={N_c}}^{\infty} \frac{\alpha^{n}}{\sqrt{n!}} \left| n \right\rangle$ \footnote{A. Gei{\ss}ler, and W. Hofstetter, arXiv:1611.10185 (submitted)}, where $N_c = 4$, leading to a negligible error in the calculated observables, which is maximal for values of $J/\Omega>0.1$, where it is on the order of a few percent ($<3\%$). The ground state is then found by comparing the resulting lattice-averaged grand canonical potentials $f = \langle H \rangle / A_{\textrm{cluster}}$, for each of the considered crystal structures. In order to allow for chequerboard order on all cluster types (even those of odd valued volume), we always simulate clusters generated by the spanning vectors ($2\cdot\mathbf{a}_1, 2\cdot\mathbf{a}_2$).

%\label{sec:f-DMFT}\textit{\color{red}Grand canonical potential within RB-DMFT} --- 
Calculation of $f$ is not straightforward within RB-DMFT, as the kinetic energy $E_{kin} = \langle H_{2BH,kin} \rangle$ is given in terms of \textit{non}-local expectation values $\langle \hat{b}^{\dagger}_{\sigma,i} \hat{b}_{\sigma,j} \rangle$, which therefore cannot directly be calculated from the self-consistent local Anderson impurity models used by RB-DMFT. But it can be shown that within the RB-DMFT self-consistency conditions, $E_{kin}$ can also be written in terms of connected local Green's functions $\mathbf{G}^C_{\sigma,i}$ and Anderson impurity hybridization functions $\mathbf{\Delta}_{\sigma,i}$, for both of which we will now give a short introduction regarding their role within DMFT. 

\subsection{Kinetic energy and connected Green's functions}

Starting from the connected normal real-space Green's function at equal times, with time ordering fixed by the infinitesimal time difference $\epsilon<0$, we have

\begin{align}
\lim_{\epsilon \rightarrow 0^{-}} G^{Cn}_{\sigma,ji}(\epsilon,0) =& -\left(\langle \hat{b}^{\dagger}_{\sigma,i} \hat{b}_{\sigma,j} \rangle - \langle \hat{b}^{\dagger}_{\sigma,i} \rangle \langle \hat{b}_{\sigma,j} \rangle\right) \label{eq:C_real-space_Greens} \\
 =& \lim_{\epsilon \rightarrow 0^{+}} \sum_{n = -\infty}^{\infty} \frac{e^{i \omega_n \epsilon}}{\beta} G^{Cn}_{\sigma,ji}(i \omega_n) \nonumber 
\end{align}
for the connected Green's functions $G^{Cn}_{\sigma,ji}(i \omega_n)$ in bosonic Matsubara frequencies. The anomalous part is accordingly given by

\begin{align}
\lim_{\epsilon \rightarrow 0^{-}} G^{Ca}_{\sigma,ji}(\epsilon,0) =& -\left(\langle \hat{b}_{\sigma,i} \hat{b}_{\sigma,j} \rangle - \langle \hat{b}_{\sigma,i} \rangle \langle \hat{b}_{\sigma,j} \rangle\right) \label{eq:C_real-space_GreensA} \\
 =& \lim_{\epsilon \rightarrow 0^{+}} \sum_{n = -\infty}^{\infty} \frac{e^{i \omega_n \epsilon}}{\beta} G^{Ca}_{\sigma,ji}(i \omega_n) \nonumber 
\end{align}
Thus expressing the total kinetic energy in terms of connected real-space Green's functions yields

\begin{align*}
E_{kin} &= - \sum_{ij\sigma} J_{ij}^{\sigma} \langle \hat{b}^{\dagger}_{\sigma,i} \hat{b}_{\sigma,j} \rangle \\ 
&= \sum_{ij\sigma} J_{ij}^{\sigma} \left( \lim_{\epsilon \rightarrow 0^{+}} \sum_{n=-\infty}^{\infty} \frac{e^{i\omega_n \epsilon}}{\beta} G^{Cn}_{\sigma,ji}(i\omega_n)  - \phi^*_{\sigma,i}\phi_{\sigma,j} \right)
\end{align*}
where $\phi_{\sigma,i} = \langle \hat{b}_{\sigma,i} \rangle$ is the local condensate order parameter of the atomic state $\sigma$ at lattice site $i$, while $J_{ij}^{\sigma}$ is the matrix of allowed hoppings in the system. This expression can be further simplified by employing both the local ((\ref{eq:local Dyson}) as in (36) from \cite{Vasic2015}) and lattice ((\ref{eq:lattice Dyson}) as in (37) from \cite{Vasic2015}) Dyson equations in Nambu notation, as regularly used within RB-DMFT. Here we suppress the state index $\sigma$, as this part of the derivation is independent of the atomic state. In Nambu notation for $n \geq 0$ the real-space lattice Green's functions are represented as $G^{Cn}_{ji}(+i \omega_n) = [\mathbf{G}^{C}_{ji}(i \omega_n)]_{11}$ and $G^{Cn}_{ji}(-i \omega_n) = [\mathbf{G}^{C}_{ji}(i \omega_n)]_{22}$, while the anomalous term is given by $G^{Ca}_{ji}(+i \omega_n) = [\mathbf{G}^{C}_{ji}(i \omega_n)]_{12} = G^{Ca}_{ji}(-i \omega_n)$ and $[\mathbf{G}^{C}_{ji}(i \omega_n)]_{12} = [\mathbf{G}^{C}_{ji}(i \omega_n)]^*_{21}$. So

\begin{align}
\mathbf{G}^{C}_i(i\omega_n)^{-1} &= i\omega_n \sigma_z + \mu\mathbf{1}_2 + \mathbf{\Delta}_i(i\omega_n) - \mathbf{\Sigma}_i(i\omega_n) \label{eq:local Dyson} \\
[\mathbf{G}^{C}(i\omega_n)^{-1}]_{ij} &= J_{ij}\mathbf{1}_2 + \delta_{ij}(i\omega_n \sigma_z + \mu\mathbf{1}_2 - \mathbf{\Sigma}_i(i\omega_n)) \label{eq:lattice Dyson}
\end{align}
where the Pauli matrix $\sigma_z$ is used due to Nambu notation. These equations are given in terms of local self energies $\mathbf{\Sigma}_i(i\omega_n)$, the Anderson impurity hybridization function $\mathbf{\Delta}_i(i\omega_n)$ and the local impurity Green's function $\mathbf{G}^{C}_i(i\omega_n) :=  [\mathbf{G}^{C}(i\omega_n)]_{ii}$ (DMFT self-consistency). Inserting $\mathbf{\Sigma}_i(i\omega_n)$ from (\ref{eq:local Dyson}) in (\ref{eq:lattice Dyson}), combined with a matrix multiplication by $\mathbf{G}^{C}(i\omega_n)$ from the right, where we are only interested in the diagonal elements, yields

\begin{align*}
&\sum_j [\mathbf{G}^{C}(i\omega_n)^{-1}]_{ij} [\mathbf{G}^{C}(i\omega_n)]_{ji} \\ = &\sum_j \left[ J_{ij}\mathbf{1}_2 - \delta_{ij}\left( \mathbf{\Delta}_i(i \omega_n)- \mathbf{G}^{C}_i(i\omega_n)^{-1} \right) \right] [\mathbf{G}^{C}(i\omega_n)]_{ji}
\end{align*}
Further using the self-consistency property of the impurity Green's function leads to the identities

\begin{align}
\sum_j J_{ij}[\mathbf{G}^{C}(i\omega_n)]_{ji} =  \mathbf{\Delta}_i(i\omega_n) \mathbf{G}_i^C(i\omega_n)
\end{align}
where only the diagonal parts are of interest to us. Considering the symmetries in Nambu notation, they allow to simplify our expression for $E_{kin}$:

\begin{align}
E_{kin} = & \frac{2}{\beta} \lim_{\epsilon \rightarrow 0^{+}} \sum_{i\sigma n\geq0} \textrm{Re}\left( \left[ \mathbf{\Delta}_{\sigma,i}(i\omega_n) \mathbf{G}^C_{\sigma,i}(i\omega_n)\right]_{11} e^{i\omega_n \epsilon} \right) \nonumber \\ & - \sum_{ij\sigma} J_{ij}^{\sigma}\phi^*_{\sigma,i}\phi_{\sigma,j} - \frac{\textrm{Tr} \left[ \mathbf{\Delta}_{\sigma,i}(0) \mathbf{G}^C_{\sigma,i}(0)\right]}{2 \beta}
\label{eq:E_kin_simplified}
\end{align}
The remaining problem is due to the cutoff imposed on the Matsubara frequencies in the numerics, which implies that the limit of equal times is not simply given by setting $\epsilon=0$. One can instead account for the cutoff by requiring that the particle number is given correctly:

\begin{align}
-  \frac{1}{\beta} \sum_n G^{C}_{R,\sigma,ii}(i\omega_n) e^{i\omega_n \epsilon} + \phi^*_{\sigma , i} \phi_{\sigma , i} \stackrel{!}{=} \langle \hat{n}^{\sigma}_i \rangle_{AIM}
\end{align}
For every site and species this yields a value of $\epsilon$ which can be used to calculate the kinetic energy in the local representation \eqref{eq:E_kin_simplified}, thereby allowing the complete calulation of the lattice-averaged grand canonical potential $f$ for each of the various crystal structures.

\subsection{Hybridization functions $\mathbf{\Delta}_{\sigma,i}$ of the effective impurity model}

The essence of RB-DMFT simulations is the mapping of a lattice model onto a set of self-consistent quantum impurity models. Primary aim of the mapping for each site is an optimal representation of the total action $S$ in terms of an effective local impurity action $S_{\textrm{eff}}$. Suppressing the pseudo-spin $\sigma$, one has

\begin{align}
S[b^*,b] = \int_0^{\beta} d\tau \left( \sum_{i} b_i^*(\tau) \frac{\partial b_i(\tau)}{\partial \tau} + H \right) = S_0 + C + \Delta S
\end{align}
with $H$ given by the model Hamiltonian (1) in terms of the boson fields $b_i(\tau)$ as functions of imaginary time $\tau$. For a given site $i\equiv0$ we also introduce both

\begin{align*}
S_0 =& \int_0^{\beta}d\tau \left(b_0^*\frac{\partial b_0}{\partial \tau}-\mu b_0^* b_0+\frac{U}{2} |b_0|^4\right) \\ \textrm{} \Delta S =& \int_0^{\beta}d\tau\sum_{\langle 0, i\rangle}\left(-t_{0i}b_0^*b_i-t_{0i}^* b_i^*b_0\right)
\end{align*}
In an approximative way, by integrating over the rest of the system, we derive the effective action $S_{\textrm{eff}}$ for a given site $i\equiv0$ via

\begin{align*}
S_{\textrm{eff}} =& S_0 + \int_0^{\beta} d\tau \sum_{\langle 0, i\rangle}-\left(t_{0i}b_0^{*} \langle b_i \rangle_C +t_{0i}^* b_0 \langle b_i^*\rangle_C\right) \\ &-\frac{1}{2}\int_0^{\beta} d\tau \int_0^{\beta} d\eta \left(b_0^*(\tau)\, b_0(\tau)\right) \mathbf{M}(\tau, \eta) \left(b_0(\eta) \,b_0^*(\eta)\right)^T
\end{align*}
where we introduce the cavity expectation value $\left\langle \cdot \right\rangle_C$, for the system where the site of interest has been removed:

\begin{align}
\langle x\rangle_C = \frac{\prod_{i\neq 0} \int\mathcal{D}b_i^* \mathcal{D}b_i x \exp(-C)}{\prod_{i\neq 0} \int\mathcal{D}b_i^* \mathcal{D}b_i\exp(-C)}
\end{align}
Then one further obtains

\begin{eqnarray*}
 M_{11}(\tau, \eta)&=&\sum_{i,j}t_{0i}t_{0j}^* \left(\langle b_i(\tau) b_j^*(\eta)\rangle_C-\langle b_i(\tau)\rangle_C  \langle b_j^*(\eta)\rangle_C \right)\\
 M_{22}(\tau, \eta)&=&\sum_{i,j}t_{0i}^*t_{0j} \left(\langle b_i^*(\tau) b_j(\eta)\rangle_C-\langle b_i^*(\tau)\rangle_C  \langle b_j(\eta)\rangle_C \right)\\
 M_{12}(\tau, \eta)&=&\sum_{i,j}t_{0i}t_{0j} \left(\langle b_i(\tau) b_j(\eta)\rangle_C-\langle b_i(\tau)\rangle_C  \langle b_j(\eta)\rangle_C \right)\\
 M_{21}(\tau, \eta)&=&\sum_{i,j}t_{0i}^*t_{0j}^* \left(\langle b_i^*(\tau) b_j^*(\eta)\rangle_C-\langle b_i^*(\tau)\rangle_C  \langle b_j^*(\eta)\rangle_C \right)
\end{eqnarray*}

On the other hand, we use exact diagonalization to solve the effective impurity model, representing the hybridization function in~\eqref{eq:local Dyson} for each site via a corresponding effective impurity bath. In this case for each site $i\equiv0$ an effective impurity model is defined as (with the internal degrees of freedom reinstated)

\begin{widetext}
\begin{align}
\begin{aligned}
H_{\mathrm{imp}}^{\mathrm{eff}} =& \sum_{\sigma} \left[-\mu \hat{b}_{\sigma,0}^{\dagger} \hat{b}_{\sigma,0}+\frac{U_{\sigma}}{2} \hat{n}^{\sigma}_{0}\left(\hat{n}^{\sigma}_{0}-1\right)-\hat{b}_{\sigma,0}^{\dagger} \left(\sum_j t_{\sigma}\langle \hat{b}_{\sigma,j}\rangle_C\right) -\hat{b}_{\sigma,0} \left(\sum_j t_{\sigma}\langle \hat{b}_{\sigma,j}\rangle_C\right)^{\dagger}\right] \\
&+\left(V_{vdW} \sum_{j\neq 0} \frac{\langle \hat{n}^e_j \rangle}{\left| \mathbf{j} \right|^6} - \Delta\right) \hat{b}_{e,0}^{\dagger} \hat{b}_{e,0} + \frac{\Omega}{2} \left( \hat{b}_{g,0}^{\dagger} \hat{b}_{e,0} + \hat{b}_{e,0}^{\dagger} \hat{b}_{g,0} \right) + U\lambda \hat{n}_0^g \hat{n}_0^e  \\
 &+\underbrace{\sum_{l,\sigma} \left( \epsilon_{l,\sigma} \hat{a}_{l,\sigma}^{\dagger} \hat{a}_{l,\sigma}+\left( V_{l,\sigma} \hat{a}^{\dagger}_{l,\sigma} \hat{b}_{\sigma,0} + V_{l,\sigma}^* \hat{a}_{l,\sigma} \hat{b}_{\sigma,0}^{\dagger} +W_{l,\sigma} \hat{a}_{l,\sigma} \hat{b}_{\sigma,0}+W_{l,\sigma}^*\hat{a}^{\dagger}_{l,\sigma} \hat{b}_{\sigma,0}^{\dagger}\right)\right)}_{H'_{\mathrm{imp}}}.
 \label{eq:anderson}
\end{aligned}
\end{align}
\end{widetext}
where $\sigma = g,e$, so $U_g=U$ and $U_e=\tilde{\lambda}U$. The hybridization function matrix $\mathbf{\Delta}_{\sigma,0}(i\omega_n)$ of this impurity model is given by all the bath terms $H'_{\mathrm{imp}}$, which include any of the bath creation and annihilation operators $\hat{a}^{\dagger}_{l,\sigma}$ and $\hat{a}_{l,\sigma}$. It has the form

\begin{figure*}[t]
 \centering
  \includegraphics[width=\textwidth]{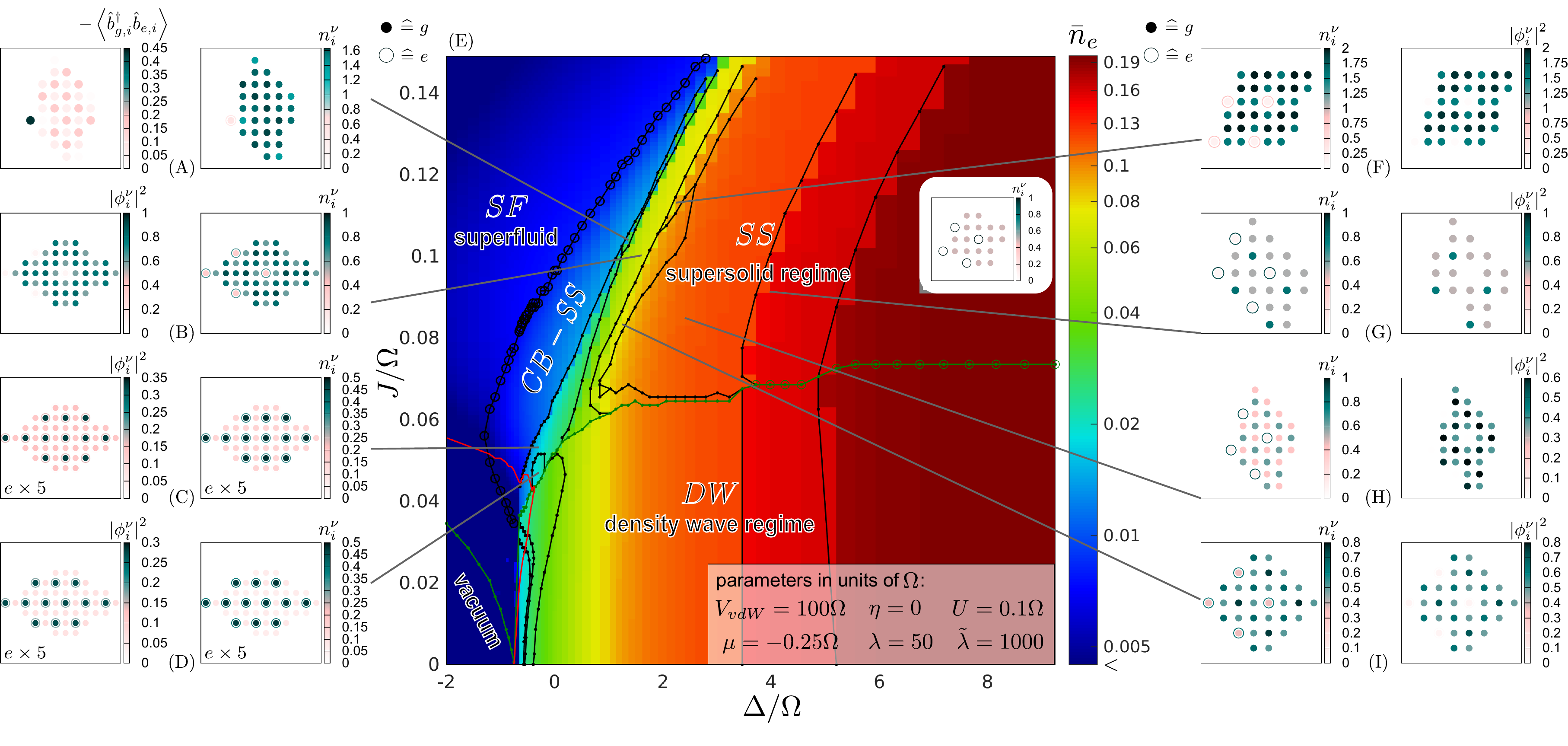}
 \caption{Colored lines in (E): Phase diagram of the two-species extended Bose-Hubbard model with vdW-interacting excited Rydberg species \eqref{eq:full-hamil}. Shown is the dependence of the average GS Rydberg fraction $\bar{n}_e$ on detuning and hopping, while the fixed parameters of the model are given in the inset. The occurrence of a finite condensate order parameter at finite $J$ is marked by the green line. Transitions between different phases of supersolid ($SS$) order above this line, as well as between density wave ($DW$) ordered phases below, are separated by black lines (circles for second order, points for first order). As it has the simplest order beyond a homogenous superfluid ($SF$), we specifically label the chequerboard supersolid ($CB-SS$) in the diagram. All DMFT results in the region between the red line and vacuum have lattice-averaged grand canonical potentials $f>0$. (E): Lattice-averaged Rydberg fraction $\bar{n}_e$, which is strongly related to the effective lifetime of Rydberg states \cite{Johnson2010}. (A-D,inset E,F-I): Depictions of the spatial distribution of respectively specified local observables. These plots correspond to different points indicated in the phase diagram (E). If mentioned in a diagram, the values for excited states are rescaled by the indicated factor.}
 \label{fig:RB-DMFT_phasediag}
\end{figure*}

\begin{align}
\mathbf{\Delta}_{\sigma,0}(i\omega_n) = \begin{pmatrix}
\Delta^{11}_{\sigma}(i\omega_n) & \Delta^{12}_{\sigma}(i\omega_n) \\
\Delta^{21}_{\sigma}(i\omega_n) & \Delta^{22}_{\sigma}(i\omega_n)
\end{pmatrix}
\end{align}
where the different matrix elements are given by

\begin{align*}
 \Delta^{11}_{\sigma}(i \omega_n) &= \sum_l\left(\frac{V_{l,\sigma}^*V_{l,\sigma}}{\epsilon_{l,\sigma}-i\omega_n}+\frac{W_{l,\sigma}^*W_{l,\sigma}}{\epsilon_l+i\omega_n}\right) \\
 \Delta^{22}_{\sigma}(i \omega_n) &= \sum_l\left(\frac{W_{l,\sigma}^*W_{l,\sigma}}{\epsilon_{l,\sigma}-i\omega_n}+\frac{V_{l,\sigma}^*V_{l,\sigma}}{\epsilon_l+i\omega_n}\right) \\
 \Delta^{12}_{\sigma}(i \omega_n) &= \sum_l\left(\frac{V_{l,\sigma}^*W^*_{l,\sigma}}{\epsilon_l-i\omega_n}+\frac{V_{l,\sigma}^*W^*_{l,\sigma}}{\epsilon_{l,\sigma}+i\omega_n}\right) \\
 \Delta^{21}_{\sigma}(i \omega_n) &= \sum_l\left(\frac{W_{l,\sigma} V_{l,\sigma}}{\epsilon_{l,\sigma}-i\omega_n}+\frac{V_{l,\sigma} W_{l,\sigma}}{\epsilon_l+i\omega_n}\right)
\end{align*}
All parameters $V,W,\epsilon$ are fixed self-consistently, so that the effective impurity-bath hybridization is the best fit of the actual impurity-lattice hybridization as extracted from \eqref{eq:local Dyson}. When self-consistency has been achieved, the relation $\mathbf{\Delta}_{\sigma,0}(i\omega_n) \equiv \mathbf{M}_{\sigma}(i\omega_n)$ holds, where $\mathbf{M}_{\sigma}(i\omega_n)$ is the representation of $\mathbf{M}_{\sigma}(\tau , \eta) = \mathbf{M}_{\sigma}(\tau - \eta)$ in terms of Matsubara frequencies.

\section{RB-DMFT phase diagram}\label{sec:f-DMFT} 
Minimizing $f$, as calculated in the described RB-DMFT scheme, with respect to the relevant crystal orders yields the ground state phase diagram shown by the lines in Fig.~\ref{fig:RB-DMFT_phasediag}(E). For selected points, we also show the spatial distribution of important local observables, such as the occupation numbers $n^{\sigma}_i$, squared condensate order parameters $\left| \phi^{\sigma}_i \right|^2$ and $\left\langle \hat{b}^{\dagger}_{g,i} \hat{b}_{e,i} \right\rangle$, the latter related to in-plane magnetization of the pseudospin. The phase boundaries are obtained from kinks (second order) and jumps (first order) in the spatially averaged observable $\bar{n}_e = \sum_i \langle \hat{n}^e_i \rangle / A^{(\mathbf{a}_1,\mathbf{a}_2)}_{cluster}$ (see Fig.~\ref{fig:RB-DMFT_phasediag}(E)), acting as order parameter. Thus we find various ground state phases, starting with the well-known homogeneous superfluid ($SF$) and the devil's staircase in the density wave ($DW$) regime at small hopping, separated by a peculiar series of supersolids. We can distinguish two distinct regimes of supersolids, dominated by either weak or strong Rydberg dressing, arising due to two competing effects. One is the melting, induced by a large hopping amplitude $J$, while the other is the crystallizing effect of the detuning $\Delta$. Since blue detuning facilitates Rydberg crystallization at higher densities, as well as a higher Rydberg fraction in general, the latter effect is easily understood. 

Traversing the phase diagram in the supersolid regime, starting at high $\Delta$ (Fig.~\ref{fig:RB-DMFT_phasediag}(E) inset) and reducing its value continuously, one first finds a series of GS supersolids with growing wavelength, until there is a sudden drop in the wavelength, accompanied by a rising Rydberg condensate and a fast drop of the Rydberg fraction for the sites with highest admixture of the Rydberg state (Fig.~\ref{fig:RB-DMFT_phasediag}(A-D)). Contrary to the devil's staircase in the $DW$ regime, the staircase in the $SS$ regime does not end in an empty or homogeneous system, but instead with short wavelength supersolids, most notably the chequerboard supersolid ($CB-SS$) (see also Fig.~\ref{fig:RB-DMFT_phasediag}(C,D)), which is the only previously predicted $SS$ phase \cite{Saha2014}. 
The competition between crystallizing and melting effects becomes especially evident in the two cases where two supersolids meet, which both have the same number of sites in their unit cells, while their spanning vectors differ (Fig.~\ref{fig:RB-DMFT_phasediag}(C,D and H,I)). There the crystallizing effect dominates for small hopping, as the excitations minimize interaction energy by maximizing their NN distances. For increased hopping the system then prefers the configuration with slightly reduced NN distances, while restoring a spatial order commensurate with the OL. Additionally the $8$-site units cells are almost degenerate, while the unit cell less favoured by $V_{vdW}$ has a transition into $SF$ at lower $J$.
Regarding the two distinct $SS$ regimes with strong and weak dressing, the narrow phase dominated by a long range order with a unit cell of 32 sites (Fig.~\ref{fig:RB-DMFT_phasediag}(A)) implies a cross-over behaviour. This phase marks the boundary between the two regimes, as it consists mostly of $CB-SS$ (with the $CB$ order strongly visible in $\langle \hat{b}^{\dagger}_{g,i} \hat{b}_{e,i} \rangle$), interspersed by a small density of strongly dressed atoms/impurities suppressing the short range $CB$ order.

Another noteworthy configuration appears in a band of width $\Delta/\Omega \approx 0.2$, starting slightly above resonance (Fig.~\ref{fig:RB-DMFT_phasediag}(B)). There the ground state condensate and the nearly Fock-state Rydberg excited atoms are spatially separated from one another, as is the case for most of the interaction dominated part of the $SS$ regime. But in addition, the excitations are aligned in a triangular lattice, while the condensate is arranged on its dual honeycomb lattice, at least as far as possible on a square lattice. 

Finally, since the effective total decay rate of excitations is directly proportional to the fraction $n_e$ of their occupation \cite{Johnson2010}, this quantity (Fig.~\ref{fig:RB-DMFT_phasediag}(E)) implies that the region with low Rydberg occupation should be most suitable for experiment. Even at detunings $\Delta > 0$, Rydberg blockade causes a value of $\bar{n}_e$, which is nearly two orders of magnitude reduced, compared to the full resonant excitation of single atoms, thus increasing the feasibility of realizing the corresponding supersolids.

In conclusion, while dressed models break down close to resonant Rydberg dressing, the combined effort of an analytically solvable ``frozen'' limit model and RB-DMFT simulations at finite hopping allows for the analysis of the rich phase diagram of~\eqref{eq:full-hamil}. In particular we find two distinct regimes of supersolid order dominated by either weak or strong dressing reminiscent of the bistable behaviour in non-itinerant dissipative systems \cite{Lee2012,Carr2013,Marcuzzi2014}. Due to our limitation to periodic systems with finite unit cells, the behavior at the cross-over remains an open question.
It should also be noted that the Rabi frequency was taken to be in the range of a few MHz, while so far realized values of hopping amplitudes only reach a fraction of this. But considering the phase diagram of the Bose-Hubbard-model, the transition to supersolid phases can be expected at strongly reduced hopping for values of $\mu$ close to zero where the assumption of low filling $\bar{n}<1$ breaks down, leaving this regime open for further research.

\begin{figure*}[t]
 \centering
  \includegraphics[width=\textwidth]{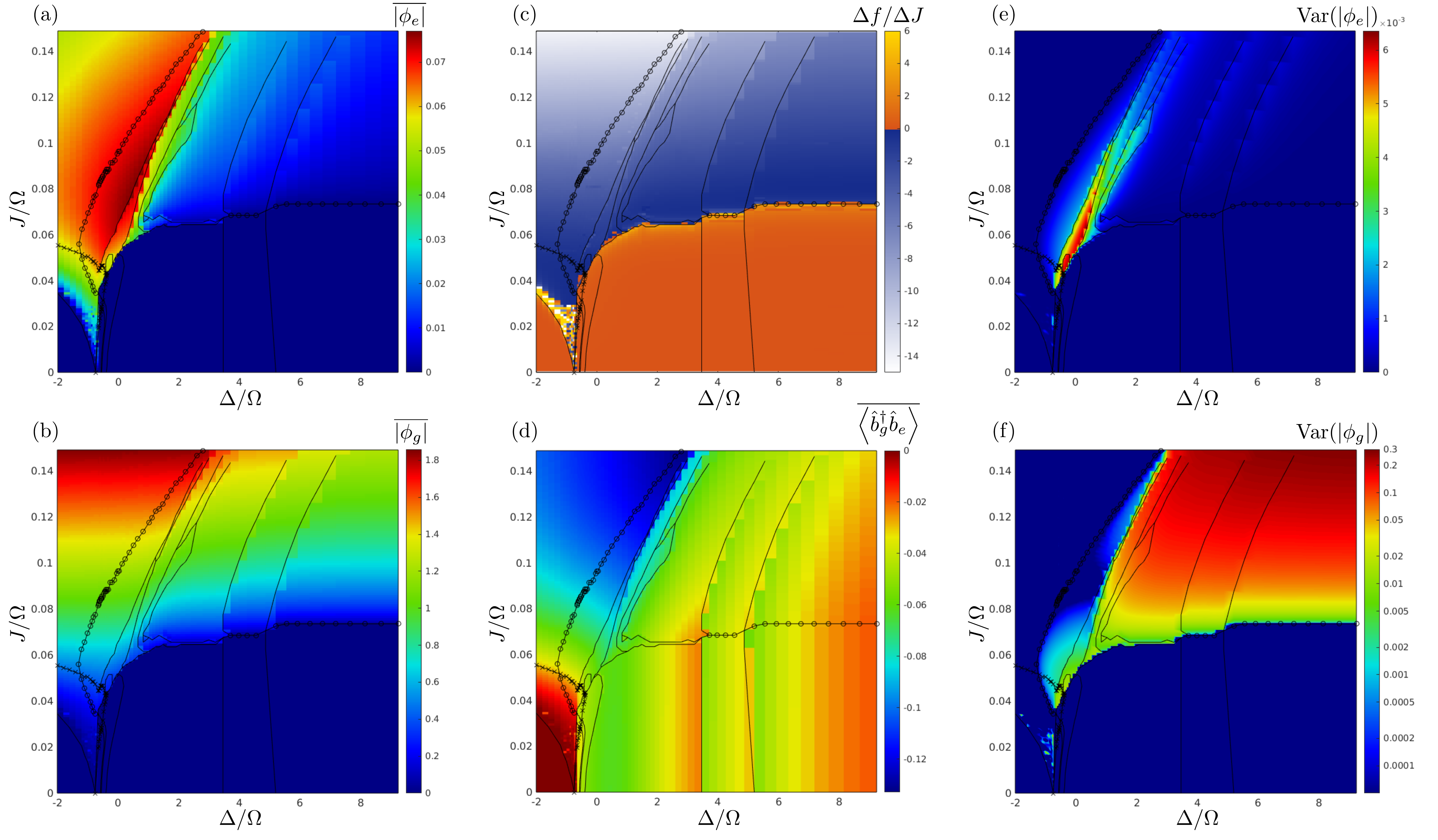}
 \caption{Here we show the different averaged order parameters one may use to distinguish the different supersolid phases as explained in the text. (a,b) Spatially averaged condensate order parameters $\overline{\left|\phi_{\sigma}\right|} = \sum_i \left|\phi^{\sigma}_i\right|/A$. Averages are normalized by the size $A$ of the system simulated within RB-DMFT. Both species have opposite but spatially constant phases, as one might expect from a dark state. (c) Difference quotient $\Delta f / \Delta J$ of the mean grand canonical potential $f$ by the hopping amplitude $J$. (d) Spatial average of the local fluctuations $\left\langle \hat{b}^{\dagger}_{g,i} \hat{b}_{e,i} \right\rangle$ induced by the Rabi term (2) of the Hamiltonian (1). A non-zero value is related to in-plane magnetization of the pseudo-spins $\sigma = \{ g, e \}$. (e,f) Spatial variance $\textrm{Var}(\left|\phi_{\sigma}\right|) = \overline{\left|\phi_{\sigma}\right|^2} - \overline{\left|\phi_{\sigma}\right|}^2$ of the condensate order parameters.
 }
 \label{fig:RB-DMFT_orderparams}
\end{figure*}

We would like to thank M. Fleischhauer, T. Nie\-der\-pr\"{u}m, H. Ott, A. Pelster, M. Weidem\"{u}ller, H. Weimer, S. Whitlock and J. Zeiher for insightful discussions. Support by the Deutsche Forschungsgemeinschaft via DFG SPP 1929 GiRyd, DFG SFB/TR 49, DFG FOR 801 and the high-performance computing center LOEWE-CSC, as well as by the DAAD via PPP Serbia (project nr. 57215082) is gratefully acknowledged. I. V. acknowledges support by the Ministry of Education, Science, and Technological Development of the Republic of Serbia under project ON 171017  and by the European Commission under H2020 project VI-SEEM, Grant No. 675121.

\appendix
\section*{Appendices}

\subsection{Further observables}
The phase boundaries for finite hopping $J$, shown in Fig.~\ref{fig:RB-DMFT_phasediag}(E), were obtained via the spatially averaged values of the local observables, which act as order parameters of the system. As can be seen in Fig.~\ref{fig:RB-DMFT_orderparams}, they exhibit either jumps or kinks at certain points in the phase diagram, allowing us to determine the phase boundaries as well as the order of the phase transitions. As the Rydberg fraction $\bar{n}_e$ exhibits the most prominent changes (see Fig.~\ref{fig:RB-DMFT_phasediag}(E)), it was used to obtain the phase boundaries between the various $SS$ and $DW$ phases.

\begin{figure*}[t]
 \centering
  \includegraphics[width=\textwidth]{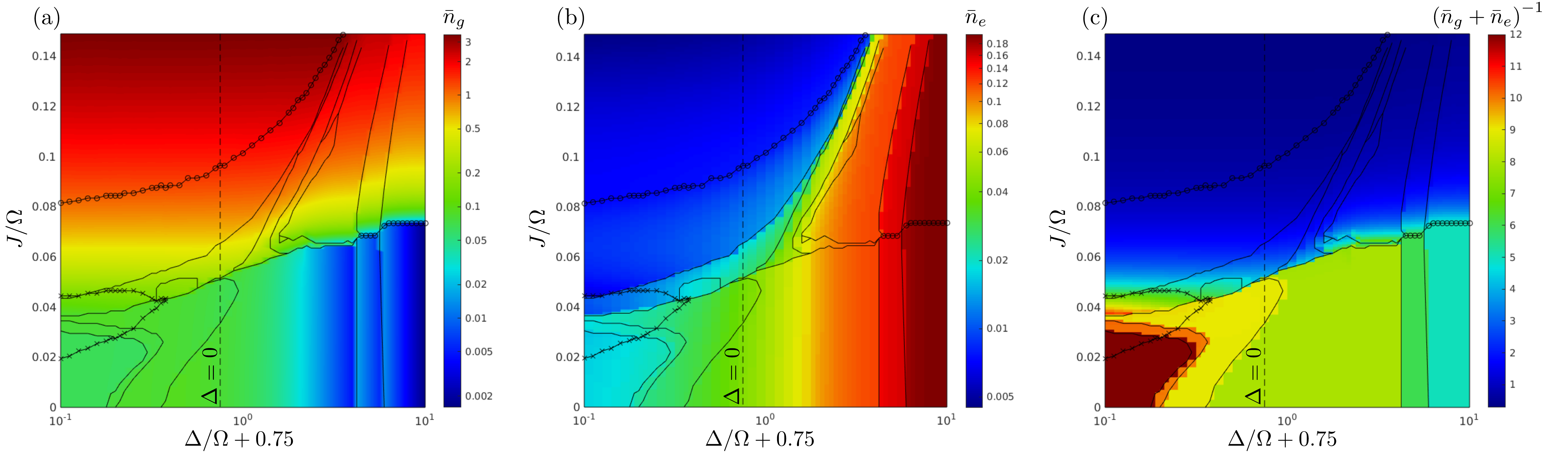}
 \caption{(a,b) Averaged occupation numbers $\bar{n}_{\sigma} = \sum_i n^{\sigma}_i/A$, where $A$ is the normalization due to the considered number of sites. (c) The inverse of the average lattice filling becomes integer in the DW regime. The values of these integers correspond to the area defined by the spanning vectors introduced earlier (compare also Fig. 1 in the main paper).
 }
 \label{fig:RB-DMFT_occupparams}
\end{figure*}

Due to the complex nature of the model \eqref{eq:full-hamil}, additional observables allow for further characterization of its ground state phases. While non-zero condensate order parameters $\phi_i^{\sigma} = \left\langle \hat{b}_{\sigma,i} \right\rangle$ determine the occurrence of a superfluid ($SF$) (see (a,b) in Fig.~\ref{fig:RB-DMFT_orderparams}), the suppression of the spatial average $\overline{\left|\phi_e\right|}$ at large $\Delta/\Omega$ is a result of the dominant interactions. The spatial variance $\textrm{Var}(\left|\phi_{\sigma}\right|) = \overline{\left|\phi_{\sigma}\right|^2} - \overline{\left|\phi_{\sigma}\right|}^2$ of the condensate order parameters (see (e,f) in Fig.~\ref{fig:RB-DMFT_orderparams}) further extends/justifies the picture of two supersolid regimes, due to the distinct behaviour at small and large $\Delta/\Omega$. A vanishing value of these variances marks the loss of crystalline order and thus the transition from $SS$ to a homogeneous $SF$. The large spatial variances in $\phi_i^{\sigma}$, on the other hand, are due to suppressed condensation on sites occupied by  atoms strongly dressed with a Rydberg state. At the crossover between the two $SS$ regimes, the observable related to the Rabi process \eqref{eq:Rabi-hamil}, $\left\langle \hat{b}^{\dagger}_{g,i} \hat{b}_{e,i} \right\rangle$ also undergoes a significant change in behaviour (see (d) Fig.~\ref{fig:RB-DMFT_orderparams}).  Regarding the transitions between the various supersolid phases, we want to point out that divergences of $\Delta f / \Delta J$ (see (c) in Fig.~\ref{fig:RB-DMFT_orderparams}) are almost absent in between $SS$ phases and remarkably also at the $SS-SF$ transition.

Note that in the region, where the ground state contribution $\bar{n}_g$ vanishes (see (a) in Fig.~\ref{fig:RB-DMFT_occupparams}), the Rydberg states become almost pure number states (compare (b,c) of Fig.~\ref{fig:RB-DMFT_occupparams}). As the corresponding property, namely that $\bar{n}_e$ nearly equals $\frac{1}{q}$, where $q$ is the area of the unit cell corresponding the inverse of the mean lattice filling at a vanishing condensate fraction, also extends into the region with a finite condensate, the Rydberg state can be understood to remain in a Fock state even for increased hopping amplitudes. Condensation then happens purely in the ground state species, which implies that the condensate part spatially separates from the long-range interacting part of the system.

\subsection{Influence of Rydberg hopping}\label{app:Rydhopp}

\begin{figure}[h]
 \centering
  \includegraphics[width=0.49\columnwidth]{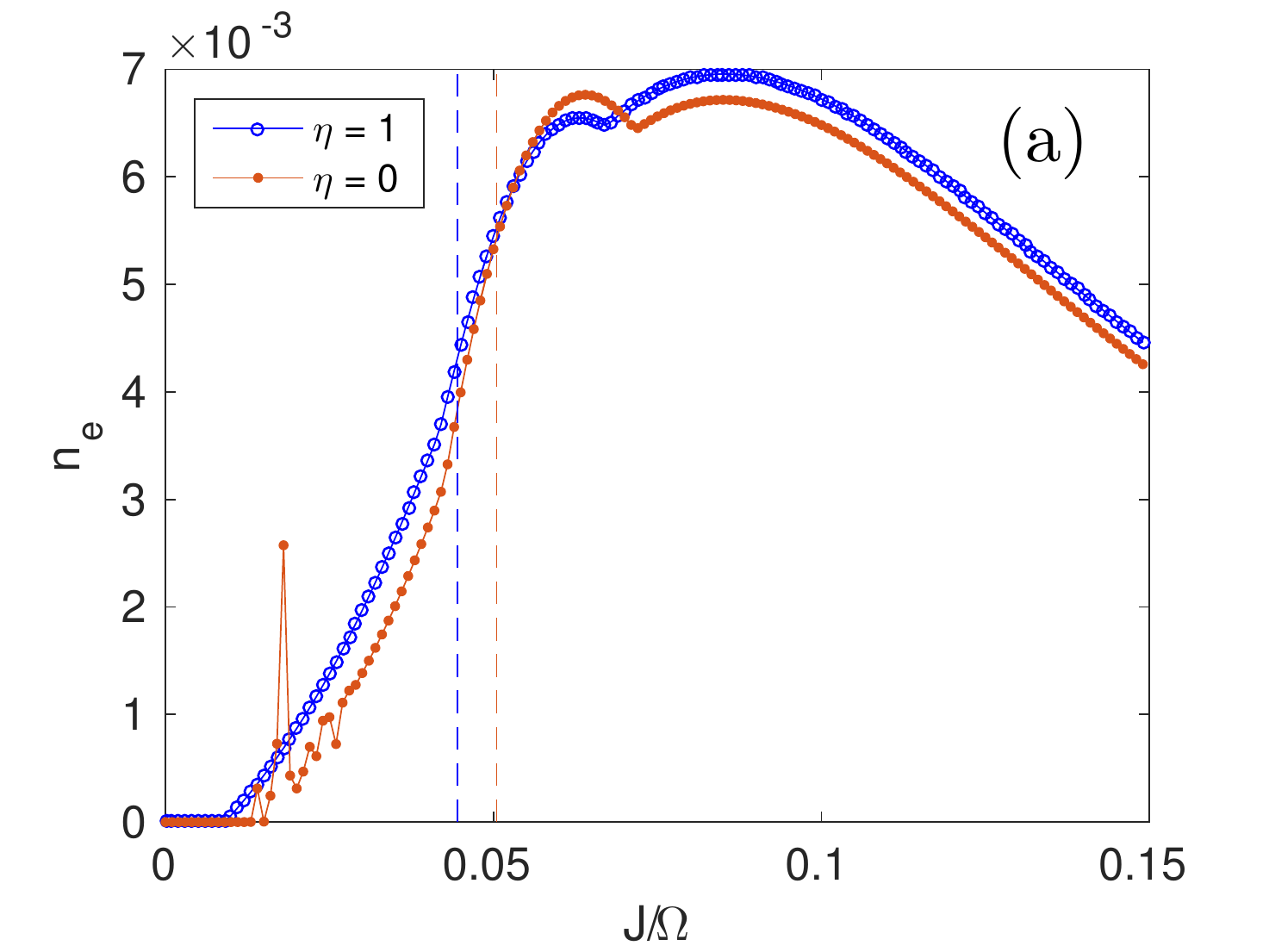}
  \includegraphics[width=0.49\columnwidth]{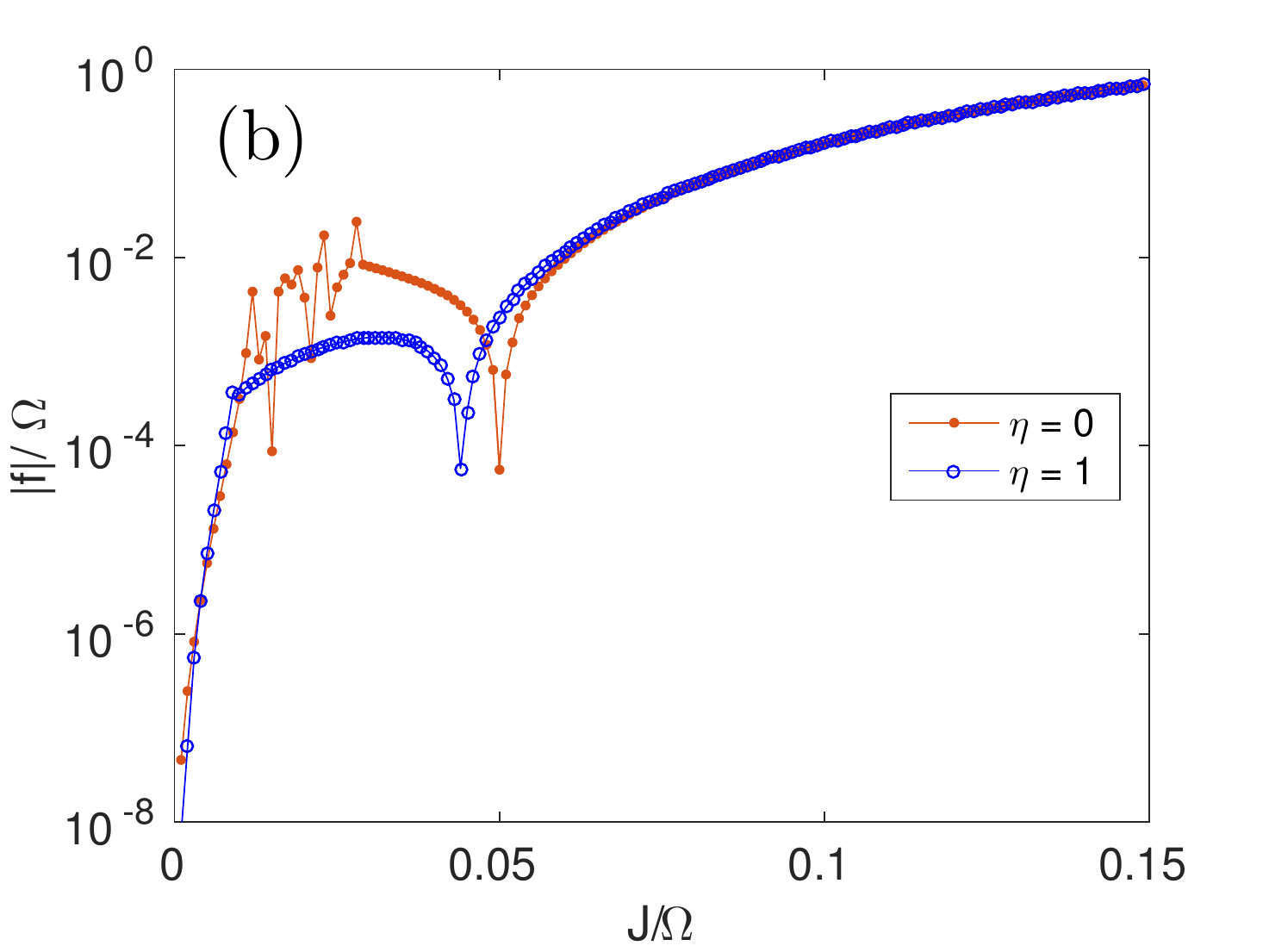}
  \includegraphics[width=0.49\columnwidth]{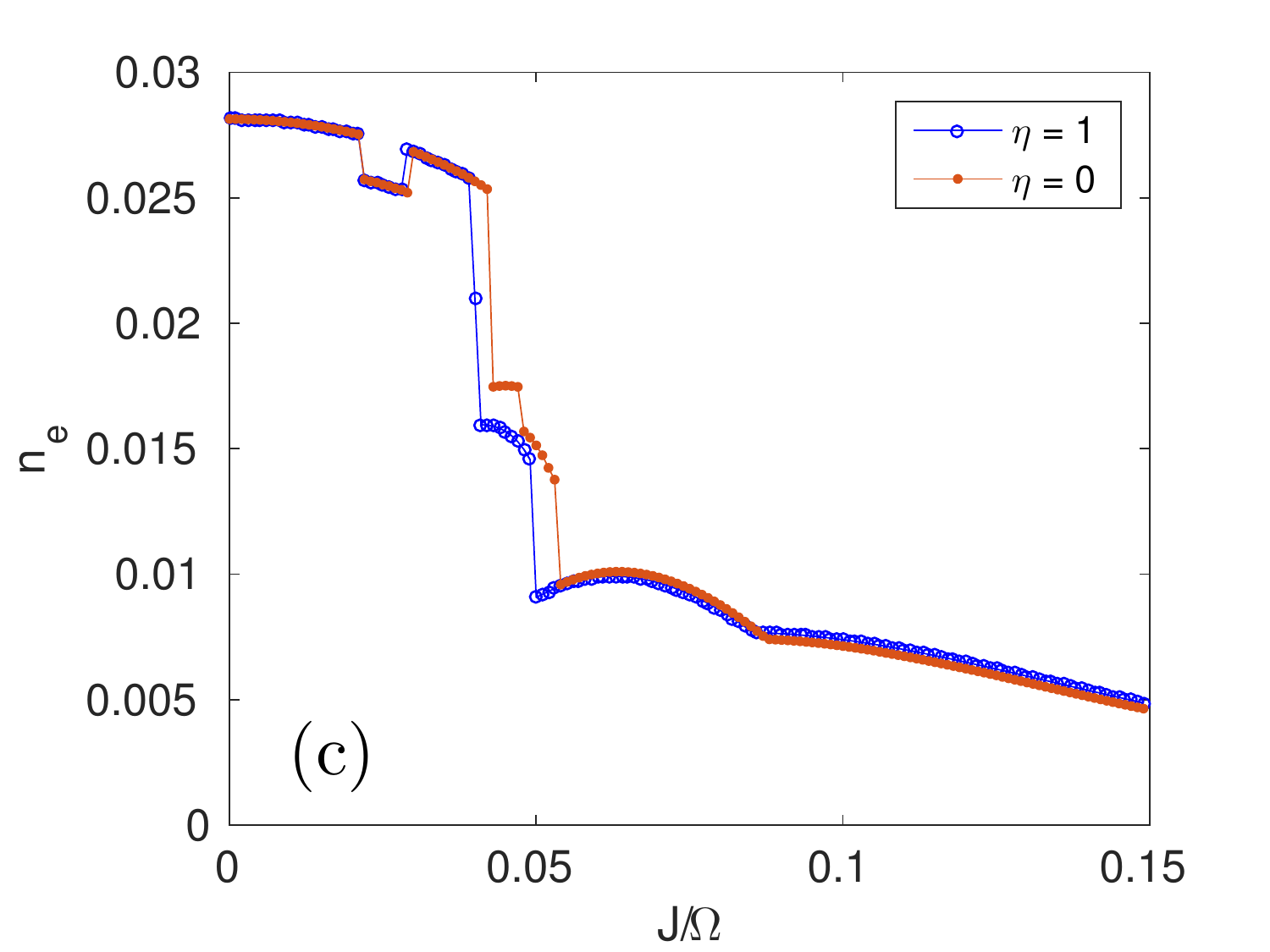}
  \includegraphics[width=0.49\columnwidth]{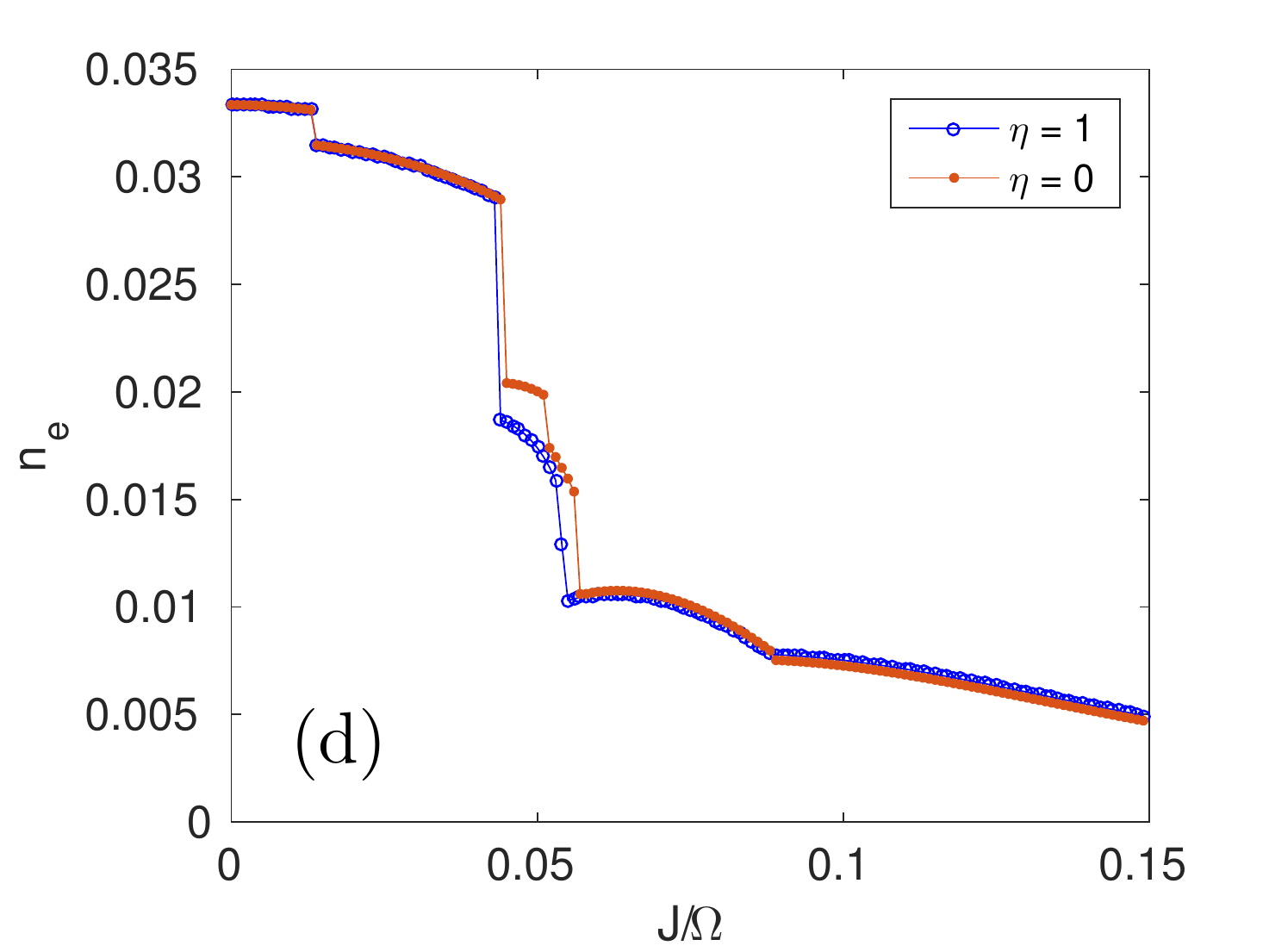}
 \caption{(a,c,d) Averaged Rydberg state occupation numbers $n_e = \bar{n}_{e}$ of (mostly) converged RB-DMFT simulations for paramters as given in the main part of the paper, except for $\eta$, which is given in the legend, while $\Delta/\Omega = -1(a);-0.415(c);-0.303(d)$. The dashed lines in (a) mark $J_c/\Omega$ where $f$ changes sign, so results at low $J/\Omega$ have a higher energy as the vacuum state $\left| n_g = 0, n_e = 0 \right\rangle$. The position of the sign change corresponds to a kink in the logarithmic plot $|f|$ as shown in (b). 
 }
 \label{fig:eta1}
\end{figure}

To further probe our assumption that we can limit itinerant behavior to the $\left|  g \right\rangle$-component, namely by setting $\eta = 0$, we also compare our results to selected simulations with $\eta = 1$. As can be seen in the comparison of the average Rydberg fraction $\bar{n}_e$, shown in Fig.~\ref{fig:eta1}, hopping of Rydberg states only has a minor influence on the phases observed in the paper. It primarily leads to changes in parameter regions, where given phases are almost degenerate. This can be seen as one of the 4-site-unit-cells vanishes for the chosen parameters, leading to one less step in Fig.~\ref{fig:eta1}(c,d). Otherwise there are only small deformations of the boundaries.

\subsection{CTS-truncation}

\begin{figure}[h]
 \centering
  \includegraphics[width=0.99\columnwidth]{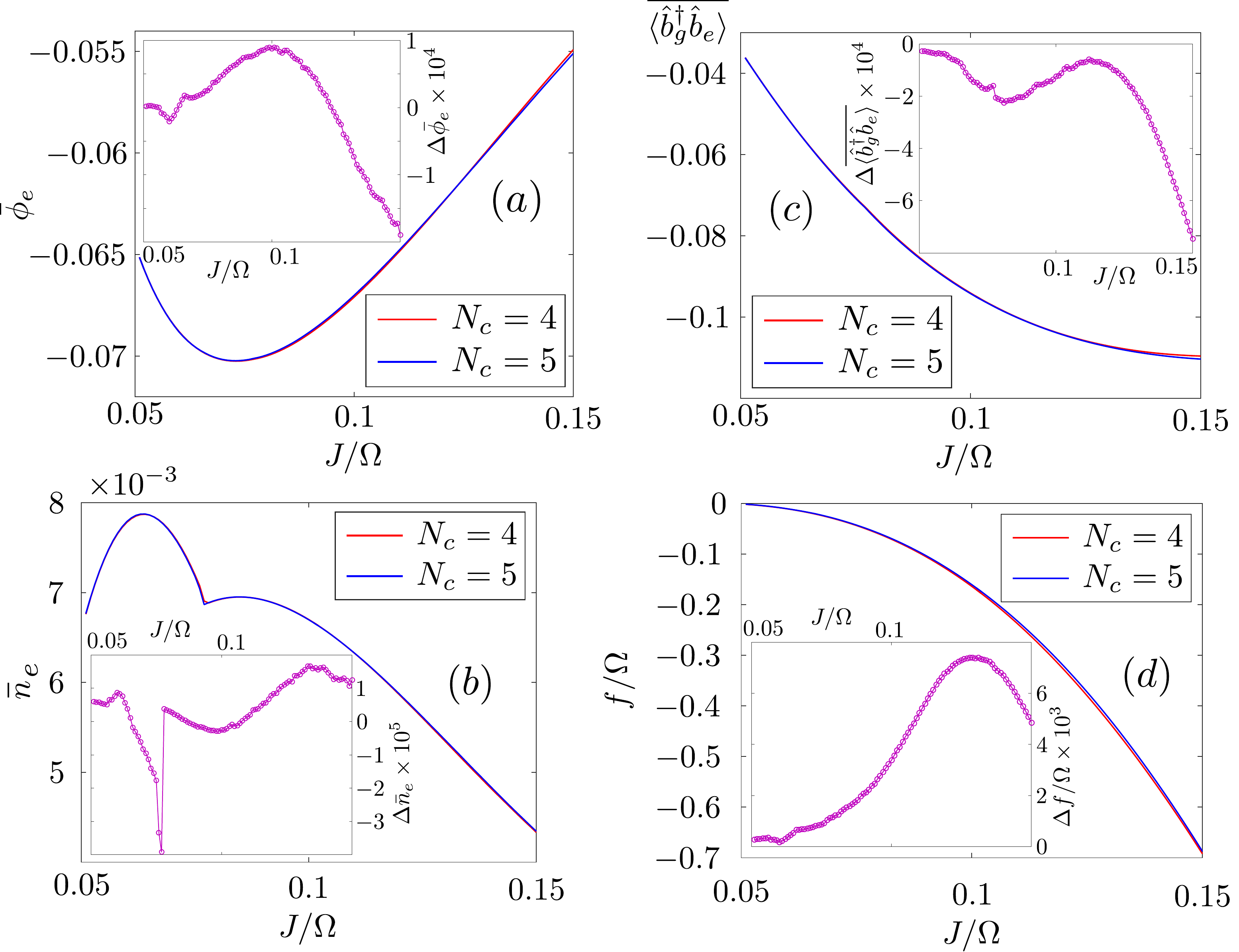}
 \caption{(a,b,c) respectively show the lattice averaged observables $\bar{\phi}_e,\bar{n}_e$ and $\overline{ \langle \hat{b}_g^{\dagger} \hat{b}_e \rangle }$ as functions of $J/\Omega$ for $\Delta/\Omega = -0.8$ with the remaining parameters as in Fig.\ref{fig:RB-DMFT_phasediag} and with a truncation scheme as given in the legends. The lattice averaged grand canonical potential $f$ is shown in (d). All the insets depict each relative deviation for the two truncation schemes $N_c = 5$ and $N_c = 4$. 
 }
 \label{fig:CTStrunc}
\end{figure}

In order to benchmark the choice of the Fock-space truncation, where we used a soft cutoff scheme, which replaces the highest Fock-state $N_c$ by the \textit{coherent-tail} state $\propto \sum_{n=N_c}^{\infty}\frac{\alpha^n}{\sqrt{n}}$ \cite{Note1}, we probed the influence of a changed truncation (i.e. changing $N_c$) on the observables and especially on the lattice-averaged grand canonical potential $f$. We did this in a parameter region where the largest deviations are expected. As the lattice filling increases above $3$ atoms per site, thus close to the used cutoff $N_c = 4$, for small $\Delta$ and large hopping (see Fig.~\ref{fig:RB-DMFT_occupparams}(a)), we chose $\Delta/\Omega = -0.8$ and $J/\Omega > 0.05$ for the benchmark. Fig.~\ref{fig:CTStrunc}(a,b,c) depicts the observables $\bar{\phi}_e,\bar{n}_e$ and $\overline{ \langle \hat{b}_g^{\dagger} \hat{b}_e \rangle }$, which have the largest deviations. As can be seen, changing $N_c$ from $4$ (as used for all the main results) to $5$ barely has any influence on these observables. The most pronounced changes appear for $J/\Omega>0.1$, with only minor numerical changes in the values of the observables, while the $SF \leftrightarrow CB-SS$ is only shifted very slightly. This can be seen from the kink in $\bar{n}_e$, as shown in Fig.~\ref{fig:CTStrunc}(b) and its inset. $f$ also experiences only minor deviations, which have a maximum around $J/\Omega \approx 1.3$, as shown in the inset of Fig.~\ref{fig:CTStrunc}(d). We therefore conclude that our results can be considered as converged with respect to the Fock-space cutoff.

\newpage

\bibliography{library}

\end{document}